\documentstyle[aps,preprint]{revtex}
\draft

\begin{document}

\title{Spin-flip and spin-wave excitations in
arbitrarily polarized quantum Hall states}
\author{Sudhansu S. Mandal \cite{email}}
\address{Department of Physics, Indian Institute of Science,
Bangalore 560 012, India \\
and\\
Condensed Matter Theory Unit, Jawaharlal Nehru Center for
Advanced Scientific Research, Jakkur, Bangalore 560 064, India}
\maketitle

\begin{abstract}
  
 We study spin-flip and spin-wave excitations for arbitrarily polarized
quantum Hall states by employing a fermionic Chern-Simons gauge theory 
in the low Zeeman energy limit. 
We show that the spin-flip correlation functions do not get renormalized 
by the fluctuations of Chern-Simons gauge field. As a consequence, the 
excitations for a given integer quantum Hall state are identical to 
fractional
quantum Hall states in the lowest Landau level having the same numerator
equal to the integer quantum Hall state.
Fully and partially polarized states  
possess only
spin-wave excitations while spin-flip excitations are possible
for all states, irrespective of their polarizations.

\end{abstract}


\section{Introduction}

By now the composite fermion model is well established in fractional
quantum Hall effect. Proposed originally by Jain \cite{jain:89} in
order to describe the fractional quantum Hall states (QHS), it was 
field theoretically developed by Lopez and Fradkin \cite{lopez:91},
and Halperin, Lee, and Read \cite{halp:93} who studied the model
near the filling fraction $\nu =1/2$. Many experiments 
\cite{du:93,will:93,kang:93,lead:94,du:94,mano:94,gold:94,ying:94,kuku:94,bayot:95,kuku:95}
have also confirmed the existence of composite fermions. Subsequently,
the composite fermion (CF) picture has been extended to describe QHS
with arbitrary polarization, by Mandal and Ravishankar \cite{ssm:96a}.
The model, which we shall call the doublet model, employs a doublet of
Chern-Simons (CS) gauge fields corresponding to two spin degrees of
freedom. Mandal and Ravishankar have shown that almost all the observed
quantum Hall states can be described by the mean field (MF) of CS gauge
fields. They have also studied the fluctuations about MF configurations.
They have further studied \cite{ssm:96b} charge density and spin density
excitations $(\delta S_z = 0)$ for these arbitrarily polarized QHS (APQHS)
using time dependent Hartree-Fock approximation (TDHFA) for the Coulomb
interaction between CF's.

In this paper, we further study the spin flip and spin wave excitations 
$(\delta S_z = \pm 1)$ for APQHS.
Stein {\it et al} ~\cite{stein:83}
were the first to observe experimentally the electron spin resonance
corresponding to spin wave excitations in odd integer QHS. 
This type of excitations is possible only for odd integer QHS;
for then, if the Lande $g$ factor is small, one spin state in the topmost
filled Landau level (LL) will be filled, leaving the other closeby spin state
unoccupied. Spin flip and spin wave excitations in these systems were
first studied theoretically by Kallin and Halperin \cite{kallin:84}.
They employed a diagrammatic approach which is equivalent
to TDHFA for the Coulomb interactions between electrons. 
Subsequently, Longo and Kallin 
~\cite{longo:93} extended the analysis of spin flip excitations to
partial filling factors. They speculated that the change in the energy of
excitations due to the Coulomb interactions
is simply the filling factor times the corresponding value for fully filled
lowest LL; the lowest energy required for spin flip excitations is the
cyclotron energy. 
In fact, we show below that the change in the energy of excitations
due to the Coulomb interactions between composite fermions for a given 
fractional filling factor in the lowest LL is equal to the corresponding
value of energy for the corresponding value of fully filled integer
state. In other words, all the QHS in the lowest LL which have the same
numerator have the same spin flip excitation energy. Further, the
lowest mode corresponding to the spin flip excitations is the effective
cyclotron energy of the CF, in consistency
with what we expect from CF picture.

We show that the CS field fluctuations do not contribute to the spin
flip correlation functions (SFCF). The latter hence may be determined
exactly (without making an expansion in wave vector). Moreover, it is
the same for all the states which have identical number of fully
filled effective LL. We find that the Coulomb interaction between CF's
produces an additional gap for spin flip excitations, in general. On
the other hand, spin wave modes are gapless as $g \rightarrow 0$, 
as required by Larmor's theorem.

The plan of the paper is as follows. In the next section, we briefly
reintroduce the doublet model and then present the formalism for 
determining SFCF corresponding to spin
density excitations (SDE) with $\delta S_z = \pm 1$. In section III,
we evaluate SFCF in the TDHFA by a diagrammatic approach for the Coulomb
interaction between the CF. In section IV, we present spin flip
and spin wave excitations for APQHS. Section V is devoted to a summary
and discussion to the paper.

\section{The formalism}

To present it briefly, consider the `doublet Lagrangian' \cite{ssm:96a}
\begin{eqnarray} 
{\cal L} &=& \psi_\uparrow^\ast {\cal D}
	 (a_\mu^+ + a_\mu^- )\psi_\uparrow 
   + \psi_\downarrow^\ast {\cal D}(
   a_\mu^+ -a_\mu^- )\psi_\downarrow
   +{\theta_+  \over 2} a_\mu^+ \epsilon^{\mu\nu \lambda }  
   \partial_\nu a_\lambda^+    \nonumber \\ 
  & &  +{\theta_-  \over 2} a_\mu^- \epsilon^{\mu\nu \lambda }  
   \partial_\nu a_\lambda^- 
    -eA_0^{\rm in}\rho 
   +e\psi^\dagger (\sigma_+ h^+ +\sigma_- h^-) \psi  \nonumber \\
    & & +{1 \over 2}\int d^3 x^\prime A_0^{\rm
   in}(x)V^{-1}(x-x^\prime)A_0^{\rm in} (x^\prime) \; ,
\label{eq1}
\end{eqnarray} 
where we have introduced additional sources $h^\pm$ which can flip spin. 
Here $\psi \equiv (\psi_ \uparrow \, ,\, \psi_ \downarrow )$
is the doublet of fermionic fields where $\uparrow (\downarrow)$
represents spin up(down).
We define ${\cal D} (a_\mu) $ as
\begin{equation}
{\cal D} (a_\mu ) = iD_0 + (1/2 m^\ast ) D_k^2 +\mu
+(g/2) \mu_{B} B \sigma
\label{eq2}
\end{equation}
with $D_\mu = \partial_\mu -ie (A_\mu +a_\mu 
+A_0^{{\rm in}} \delta_{\mu 0})$, where $A_\mu $ is
the external electromagnetic field. 
$a_\mu^\pm $ are the CS gauge fields which interact in phase with spin up
particles while they interact out of phase with spin down particles. The
field $A_0^{{\rm in}}$ is internal scalar potential. The mean particle
density $\rho$ is held fixed by the chemical potential $\mu $. 
$m^\ast $ is the effective mass of the particles.
The Zeeman term includes applied magnetic field \cite{fnote}. 
$\mu_B$ is the Bohr magneton and $\sigma = +1 (-1)$ for spin-up (-down)
particles. The potential for interactions between CF
is considered to be Coulombic, {\it i.e.},
$V(r) =e^2 /\epsilon r$, where $\epsilon$ is the background
dielectric constant of the system. 
$\sigma_\pm = \frac{1}{2}(\sigma_x
\pm i\sigma_y)$ are the spin raising and lowering operators
respectively.

\subsection{Mean field results (a brief resume)}

As in Ref.~\cite{ssm:96a}, we parametrize $\theta_\pm = (e^2 / 2\pi )
(1/s_\pm )$ and set $s_- = 0$ and $s_+ = 2s$ (even integer). 
For this choice, the field $a_\mu^-$ provides a vanishing 
mean magnetic field $\langle b^- \rangle $ and it essentially decouples.
The CF picture is enforced by the choice $s_+ =2s$. Thus, in the 
MF ansatz, the CS magnetic field produced by the particle is 
$\langle b^+ \rangle = -e \rho /\theta_+ $.
The  mean magnetic field experienced by the particles, irrespective of their
spin, is given by $\bar{B}=B+\langle b^+ \rangle $. Let $p_
\uparrow (p_ \downarrow )$ be the number of LL, which are formed
by effective field $\bar{B}$,
filled by spin up (down) particles. This leads to
the actual filling fraction and spin density \cite{ssm:96a}
\begin{equation}
\nu ={p_ \uparrow +p_ \downarrow \over 2s(p_ \uparrow +p_
\downarrow )+1}\; ,\; \Delta\rho = \rho \left( {p_ \uparrow -p_
\downarrow \over p_ \uparrow +p_ \downarrow } \right) \; .
\label{eq3}
\end{equation}
Note that $p_ \uparrow $ and $p_ \downarrow $ can be negative
integers as well in which case $\bar{B}$ is antiparallel to $B$.
The effective cyclotron frequency $\bar{\omega}_c$ is related to
the actual cyclotron frequency $\omega_c = {e \over m^\ast}B$ by
$\omega_c =\bar{\omega}_c [2s(p_ \uparrow +p_ \downarrow )+1]$.
For unpolarized QHS, $p_ \uparrow =p_ \downarrow =p$ (say) and
therefore the states with filling fraction $\nu =2p/(4sp+1)$ are
spin unpolarized in the limit of small Zeeman energy. In this
limit, $p_ \uparrow =p_ \downarrow +1$ for partially polarized
states with $\Delta\rho /\rho =1/(p_ \uparrow +p_ \downarrow )$.
Fully polarized Laughlin states are obtained for $p_ \uparrow
=1$, $p_ \downarrow =0$. The integer QHS's correspond to the
choice $s=0$ ({\it i.e.}, $\theta_+ =\infty$), in which case mean CS
magnetic field is zero.

\subsection{Effective action}

Employing the above MF ansatz along with vanishing mean electric fields
$\langle {\bf e}^\pm \rangle $ and $\langle A_0^{{\rm in}} \rangle 
= 0$, we then evaluate the one-loop
effective action for the gauge
fields and probes to be
\begin{eqnarray} 
S_{\rm eff} &=&
 -{1 \over 2}\int d^3 x \int d^3 x^\prime \left[
 (a_\mu^+ +a_\mu^- +A_0^{\rm in}\delta_{\mu 0})(x)\,
 \Pi^{\mu\nu}_ \uparrow (x, x^\prime)\,
 (a_\nu^+ +a_\nu^- +A_0^{\rm in}\delta_{\nu 0}) (x^\prime)
 \right. \nonumber \\ & &
 +(a_\mu^+ -a_\mu^- +A_0^{\rm in}\delta_{\mu 0})(x)\,
 \Pi^{\mu\nu}_ \downarrow (x, x^\prime)\,
 (a_\nu^+ +a_\nu^- +A_0^{\rm in}\delta_{\nu 0}) (x^\prime)
 \nonumber \\ & &
 -A_0^{\rm in}(x)V^{-1}(x-x^\prime)A_0^{\rm in}(x^\prime)
 +2h^a (x) \Gamma_ \uparrow^{a\mu} (x, x^\prime) (a_\mu^+ +
 a_\mu^- +A_0^{\rm in}\delta_{\mu 0}) (x^\prime)
 \nonumber \\ & &
 \left. +2h^a (x) \Gamma_ \downarrow^{a\mu}(x, x^\prime) (a_\mu^+ -
 a_\mu^- +A_0^{\rm in}\delta_{\mu 0}) (x^\prime)
 +h_a (x)\chi^{ab}(x, x^\prime)h_b (x^\prime) 
 \right] \nonumber \\ & &
 +\frac{1}{2} \int d^3x \left\{
 \theta_+ \epsilon^{\mu\nu\lambda}a_\mu^+ \partial_\nu
 a_\lambda^+
 +\theta_- \epsilon^{\mu\nu\lambda}a_\mu^- \partial_\nu
 a_\lambda^- \right\}    \; \;\; ; \; \; \;  a, \, b = \pm    \, \; .
 \label{eq4}
\end{eqnarray} 
Here $a_\mu^\pm $ and $A_0^{\mbox{in}}$ are fluctuating parts of
the corresponding gauge fields. 
Note that we have kept $a_\mu^-$ field in the effective action
for the sake of completeness, although it decouples for the
states (\ref{eq3}) at hand. At the end of the calculation, however,
the limit $\theta_- = \infty $ has to be taken.
The correlation functions $\Pi^{\mu\nu}_r (x, x^\prime)$,
$\chi^{ab} (x,x^\prime)$, and $\Gamma^{a\mu}_r (x,x^\prime)$
have to be evaluated at the prescribed MF configuration. Their
explicit forms are as follows:
\begin{eqnarray}
\Pi^{\mu\nu}_r (x,x^\prime) &=& -i \left\langle j^\mu_r (x)
j^\nu_r (x^\prime \right \rangle_C - \left \langle \frac{\delta
j^\mu_r (x)}{\delta {\cal A}_\nu (x^\prime)} \right \rangle
\; , \label{eq5} \\ 
\chi^{ab} (x,x^\prime) & =& -i \left \langle j^a (x) j^b
(x^\prime) \right \rangle_C   \; , \label{eq6}  \\
\Gamma^{a\mu}_r (x,x^\prime) &=& -i \left \langle j^a (x)
j^\mu_r (x^\prime) \right \rangle_C \; . \label{eq7}
\end{eqnarray}
Here $ \langle \cdots \rangle $ represents the expectation value
in the ground state of the system. $\langle \cdots \rangle_C$
corresponds to the connected diagrams which contribute to the
expectation value. ${\cal A}_\mu$ represents the sum of 
all the gauge fields.
The current operators in Eqs.~(\ref{eq5}--\ref{eq7}) are given by
\begin{eqnarray}
j^0_r (x) &=& e\psi^\ast_r \psi_r \; , \label{eq8} \\
j^k_r (x) &=& e\frac{e}{2m^\ast} \left[ \psi^\ast_r D^k
\psi_r - \left( D^{k\ast} \psi_r^\ast \right) \psi_r \right]
\; , \label{eq9} \\
j^a (x) &=& e \psi^\dagger \sigma_a \psi  \label{eq10} \; .
\end{eqnarray}

It is easy to see that
\begin{equation}
\Gamma^{a\mu}_r (x,x^\prime) \equiv 0 \; , \label{eq11}
\end{equation} 
since $\langle \sigma_a \rangle  \equiv 0$. Therefore, the terms
in Eq.~(\ref{eq4}) corresponding to the probe $h_a$ completely
decouple from the fluctuation of the gauge fields. In other
words, gauge field fluctuations do {\em not } change the correlation
$\chi^{ab} (x,x^\prime)$. Further, $\chi^{++} (x,x^\prime) =
\chi^{- -}(x,x^\prime) \equiv 0$ since $\langle \sigma_+^2
\rangle = \langle \sigma_-^2 \rangle \equiv 0$. We thus obtain
\begin{equation}
S_{{\rm eff}}[h^+\, ,\, h^-] = -\frac{1}{2} \int d^3 x\int d^3
x^\prime \left[ h^+ (x) \chi^{+ -}(x,x^\prime) h^- (x^\prime) +
h^- (x) \chi^{- +} (x,x^\prime) h^+ (x^\prime) \right]
\label{eq12}  \; . 
\end{equation}
$\chi^{+ -} (x,x^\prime) \equiv  \delta^2 S_{{\rm eff}} /\delta
h^+ (x)\delta h^- (x^\prime)$ is the spin correlation function
where a particle of up spin is destroyed at the point $x^\prime$
and a particle of spin down is created at the point $x$. In
other words, this is the correlation function for producing a
spin up quasihole at the point $x^\prime$ and a spin down
quasiparticle at the point $x$. Similarly, $\chi^{- +}
(x,x^\prime)$ represents the spin correlation function for
producing a spin down quasihole at the point $x^\prime$ and a
spin up quasiparticle at the point $x$. These correlation
functions are the response functions for 
spin density excitation
$(\delta S_z =\pm 1)$ which we shall evaluate below.

\section{Response functions}

In terms of the single particle Green function
\begin{equation}
G (x,x^\prime) =-i \langle T \psi (x) \psi^\dagger
(x^\prime) \rangle \; ,
\label{eq13}
\end{equation}
the linear response functions can be written as
\begin{eqnarray}
\chi^{+ -} (x,\, x^\prime) &=& ie^2 \, {\rm Tr} \, \left[
\sigma_+ G (x,\, x^\prime) \sigma_- G (x^\prime ,\, x)
\right]  \; , \label{eq14}  \\
\chi^{- +} (x,\, x^\prime) &=& ie^2 \, {\rm Tr} \, \left[
\sigma_- G (x,\, x^\prime) \sigma_+ G (x^\prime ,\, x)
\right]  \; . \label{eq15}  
\end{eqnarray}
Let $G_0$ be the single particle Green's function evaluated by
switching off the Coulomb interaction. Then $\chi_0^{+ -}$ and
$\chi_0^{- +}$ that emerge are in pure random phase approximation (RPA). 
We shall go beyond RPA
and determine $\chi^{+ -}$ and $\chi^{- +}$ in the TDHFA.

We now determine the response functions (in momentum space) in
the TDHFA as a generalization to the RPA.   
Here we employ the diagrammatic approach which was developed by
Kallin and Halperin \cite{kallin:84}.
Recall that in TDHFA,
the Green's functions are Hartree Fock Green's functions 
and incorporate self energy due to the Coulomb interactions.
In this approximation, 
only those diagrams with one exciton
present at a time are considered. In other words, the Coulomb
energy $e^2/ \epsilon l_0$ is taken to be 
smaller than $\bar{\omega}_c$,
where $l_0 =(e\bar{B})^{-1/2}$ is the effective magnetic
length of the system and $ \epsilon $ is the background
dielectric constant.
This assumption is, therefore, not valid at or near $\nu =1/2s$.
In other words, the condition on the
validity of our assumption is $ e^{3/2}m^\ast / \epsilon \ll
\sqrt{\bar{B}}$.

The TDHFA response functions are determined in the appendix A. They
are
\begin{eqnarray}
\chi^{+ -} (\omega , {\bf q}) &= &\frac{e^2}{4\pi} {\bf q}^2
e^{- \bar{{\bf q}}^2 }  \nonumber \\
& & \times \left[ 
\sum_{n_2 <p_ \uparrow} \sum_{n_1 \geq p_ \downarrow }\left(
\frac{n_2 !}{n_1 !} \right) \frac{ ( \bar{{\bf q}}^2 )^{n_1
-n_2 -1} \left\{ L_{n_2}^{n_1 -n_2} (\bar{{\bf q}}^2)
\right\}^2}{ \omega -(\epsilon_{n_1}^\downarrow -
\epsilon_{n_2}^\uparrow ) -E_{n_1 n_2}^{\downarrow \uparrow }
+\tilde{V}^{(1)}_{n_1n_2n_2n_1} (q) +i \eta } \right. 
\nonumber \\
& &- \left. \sum_{n_2 <p_ \downarrow}\sum_{n_1 \geq p_ \uparrow }\left(
\frac{n_2 !}{n_1 !} \right) 
\frac{ ( \bar{{\bf q}}^2 )^{n_1
-n_2 -1} \left\{ L_{n_2}^{n_1 -n_2} (\bar{{\bf q}}^2)
\right\}^2 }{ \omega +(\epsilon_{n_1}^\uparrow -
\epsilon_{n_2}^\downarrow )+E_{n_1 n_2}^{\uparrow \downarrow }
-\tilde{V}^{(1)}_{n_1n_2n_2n_1} (q) -i \eta }\right]
\; , \nonumber \\
& & \label{eq16} \\
\chi^{- +} (\omega , {\bf q}) & = & \frac{e^2}{4\pi} {\bf q}^2
e^{- \bar{{\bf q}}^2 } \nonumber \\
& &  \times \left[ 
\sum_{n_2 <p_ \downarrow} \sum_{n_1 \geq p_ \uparrow }\left(
\frac{n_2 !}{n_1 !} \right) \frac{ ( \bar{{\bf q}}^2 )^{n_1
-n_2 -1} \left\{ L_{n_2}^{n_1 -n_2} (\bar{{\bf q}}^2)
\right\}^2 }{\omega -(\epsilon_{n_1}^\uparrow -
\epsilon_{n_2}^\downarrow ) -E_{n_1 n_2}^{\uparrow \downarrow }
+\tilde{V}^{(1)}_{n_1n_2n_2n_1} (q) +i \eta } \right. 
\nonumber \\
& & -\left. \sum_{n_2 <p_ \uparrow}\sum_{n_1 \geq p_ \downarrow }\left(
\frac{n_2 !}{n_1 !} \right) 
\frac{ ( \bar{{\bf q}}^2 )^{n_1
-n_2 -1} \left\{ L_{n_2}^{n_1 -n_2} (\bar{{\bf q}}^2)
\right\}^2 }{\omega +(\epsilon_{n_1}^\downarrow -
\epsilon_{n_2}^\uparrow )+E_{n_1 n_2}^{\downarrow \uparrow }
-\tilde{V}^{(1)}_{n_1n_2n_2n_1} (q) -i \eta } \right]
\; , \nonumber \\
& &  \label{eq17} 
\end{eqnarray}
where $\bar{{\bf q}}^2 = {\bf q}^2l_0^2 /2 $ and 
$n_1$, $n_2$ represent the indices of the LL formed by
effective magnetic field $\bar{B}$. Here $\epsilon_n^r =
(n+1/2) \bar{\omega}_c -(1/2)g\mu_B B \sigma $
is the energy of $n$ th LL with spin index $r$.
$\sigma = +1 (-1)$ for up (down) states. $E_{n_1n_2}^{rr^\prime}
= \Sigma_{n_1}^r -\Sigma_{n_2}^{r^\prime}$ represents the
exchange energy, {\it i.e.}, the difference in self energy of
particles in two different LL with unequal spin indices.
$\Sigma_n^r$ is independent of momentum $q$, and is given by
Eq.~(\ref{eqb2}).
The interaction between an excited particle and hole ({\it i.e.}, the
ladder diagrams) is represented by the matrix element
$\tilde{V}^{(1)}_{n_1n_2n_2n_1} (q)$ which is expressed in
Eq.~(\ref{eqb12}).
We note that the bubble diagrams in which a particle-hole
pair recombines to form another particle-hole pair do not
contribute to the response functions considered here. This is
because particle and hole possess different spin.

In obtaining Eqs.~(\ref{eq16}) and (\ref{eq17}), we have
included all the contributions up to order $e^2 / \epsilon l_0$
and so the form factors are essentially exact in the strong
effective magnetic field limit. In the absence of Coulomb
interaction, $\Sigma_n^r$ and $\tilde{V}^{(1)}_{n_1n_2n_2n_1}
(q)$ are zero and the response functions acquire their
pure RPA form. We shall see below that the coulomb interaction
between composite fermions plays a decisive role in the
excitations considered that we are interested in.

\section{excitations}

As we have seen above, the response functions $\chi^{+ -}$ and
$\chi^{- +}$ remain unchanged by the fluctuation of the gauge
fields. They depend only on the mean effective magnetic field.
Therefore, the corresponding SDE depend solely on the effective
magnetic length $l_0$ which is related to the actual magnetic length 
of the system $l =(eB)^{-1/2}$,  {\it via} $l_0 = (\vert p_ \uparrow +p_
\downarrow \vert /\nu )^{1/2}l$. The excitations for the
fractional states with $\nu =
\vert p_ \uparrow +p_ \downarrow \vert / (2s \vert p_ \uparrow
+p_ \downarrow \vert \pm 1)$ are equivalent to
that of integer states with $\nu = \vert p_ \uparrow 
+p_ \downarrow \vert $,
since the states have same $l_0$.

The dispersion relation for the excitation of a spin down
particle and a spin up hole is obtained from Eq.~(\ref{eq16}) as
\begin{equation}
\omega = (n_1 -n_2)\bar{\omega}_c +g\mu_B B
+E_{n_1 n_2}^{\downarrow \uparrow
}-\tilde{V}^{(1)}_{n_1n_2n_2n_1}(q) \; ,
\label {eq18}
\end{equation}
with $n_2 < p_ \uparrow $ and $n_1 \geq p_ \downarrow $. In
these SDE, the $z$ component of the spin changes as $\delta S_z
=-1$. Similarly the SDE with a spin up particle and a spin up
hole ($\delta S_z =+1$) have dispersion relation (see
Eq.~(\ref{eq17}))
\begin{equation}
\omega = (n_1 -n_2)\bar{\omega}_c -g\mu_B B
+E_{n_1 n_2}^{\uparrow \downarrow
}-\tilde{V}^{(1)}_{n_1n_2n_2n_1}(q) \; ,
\label {eq19}
\end{equation}
with $n_2 < p_ \downarrow $ and $n_1 \geq p_ \uparrow $.

\subsection{Spin-flip excitations}

In spin flip excitations, a particle changes the Landau
level as well as flipping its spin. The excitation modes can have energy
different from $(n_1 -n_2)\bar{\omega}_c +g\mu_B
B (\delta S_z)$ at $q=0$, since the Coulombic
interaction changes the gap energy, in general. There are
three types of ground state to consider: (i) fully polarized
states $(p_ \uparrow =1,\, p_ \downarrow =0)$; (ii) unpolarized
states $(p_ \uparrow =p_ \downarrow )$; and (iii) partially
polarized states $(p_ \uparrow =p_ \downarrow +1)$. We 
discuss the spin-flip excitations of each of these ground states.

Case--I: For fully polarized states, the dispersion relation,
corresponding to spin down particle and spin up hole
excitations, from Eq.~(\ref{eq18}), is given by
\begin{eqnarray}
\omega_m -m\bar{\omega}_c -g\mu_B B &=& \Delta
E_m (q) \nonumber \\
&=& E_{m0}^{\downarrow \uparrow }- \tilde{V}^{(1)}_{m00m} (q) 
\; , \label{eq20}
\end{eqnarray}
where $m$ is an integer. The changes in energy, due to the Coulomb
interaction between the fermions, corresponding to the two
lowest modes with $m=1$ and $m=2$, are respectively obtained as
\begin{eqnarray}
\Delta E_1 (q) &=& 
\frac{e^2}{\epsilon l_0} \frac{1}{2} \sqrt{\frac{\pi}{2}} 
\left\{ 2- e^{- \bar{{\bf q}}^2 /2}
\left[ (1+ \bar{{\bf q}}^2) I_0 \left( \frac{\bar{{\bf
q}}^2}{2} \right) - \bar{{\bf q}}^2 I_1 \left(
\frac{\bar{{\bf q}}^2}{2} \right) \right] \right\} 
\; , \label{eq21} \\
\Delta E_2 (q) &=& \frac{e^2}{\epsilon l_0} \frac{1}{8}
\sqrt{\frac{\pi}{2}} \left\{ 8- e^{- \bar{{\bf q}}^2 /2}
\left[ (3+ 2\bar{{\bf q}}^2 + 2 \bar{{\bf q}}^4)
I_0 \left( \frac{\bar{{\bf
q}}^2}{2} \right) \right. \right. \nonumber \\
& & \;\;\;\;\;\;\;\left. \left. - (4\bar{{\bf q}}^2 +2 \bar{{\bf
q}}^4) I_1 \left(
\frac{\bar{{\bf q}}^2}{2} \right) \right] \right\} 
\; . \label{eq22}
\end{eqnarray}
Here $I_0$ and $I_1$ are the modified Bessel's functions whose
integral representations are given by
\begin{eqnarray}
I_0 (z) &=& \frac{1}{4\pi} \int_0^{4\pi} e^{-z\cos \theta } 
 d \theta  \; , \\
\frac{1}{z}I_1 (z) &=& \frac{1}{4\pi} \int_0^{4\pi} e^{-z\cos \theta } 
 \sin^2 \theta \, d \theta  \; .
 \end{eqnarray}
These two dispersion relations are shown in Fig.~1. 
For very low $ql_0$, these energies behave as 
\begin{eqnarray}
\Delta E_1(q)& \approx &\frac{e^2}{\epsilon l_0} \frac{1}{2}
\sqrt{\frac{\pi}{2}} \left[ 1- \frac{\bar{{\bf q}}^2}{2} \right]
\; , \label{eq25} \\
\Delta E_2(q) &\approx& \frac{e^2}{\epsilon l_0} \frac{1}{8}
\sqrt{\frac{\pi}{2}} \left[ 5- \frac{\bar{{\bf q}}^2}{2} \right]
\;  . \label{eq26} 
\end{eqnarray}
Thus the Coulomb interaction has changed the gap energy for
these spin-flip excitations. At low momentum, the exchange in
self energy dominates over the interaction energy between
excited particle and hole pair. On the other hand, their
contributions reverse for higher momentum. 
Thus the dispersion relations show a minima in their spectra
(see Fig.~1).
There is no mode corresponding to the dispersion relation
(\ref{eq19}), since the population of down spins is zero in the
fully polarized ground state.

Note that the above dispersion relations of the spin-flip
excitations hold for the Laughlin
states with $\nu = \frac{1}{2s+1}$ because they all have the
same $l_0$. 
All these states acquire non vanishing gap energies due to the
Coulomb interaction, contrary to the result of Kallin and
Halperin \cite{kallin:84}
for $\nu =1$ state $(s=0)$. The reason for the mismatch between the
two results is that they have calculated exchange self energy
wrongly. This drawback was corrected by Longo and Kallin
\cite{longo:93},
who have further calculated
dispersion relations using the single mode approximation (SMA) to
obtain finite gap energies due to the Coulomb interaction in
partially filled Landau levels.
As we have seen, the
present model allows for spin-flip excitation at $
\bar{\omega}_c +g\mu_B B$, while the lowest
spin-flip mode in the SMA is at $\omega_c +g\mu_B B$.

Case--II: For unpolarized QHS, equal number of LL filled by
particles of up and down spins. Consider the simplest case $\vert 
p_ \uparrow \vert = \vert p_ \downarrow \vert =1$. The
dispersion relations for the spin flip excitations with $\delta
S_z = \mp 1$ are respectively given by
\begin{eqnarray}
\omega_m -m \bar{\omega}_c -g\mu_B B &=& \Delta
E_m^-  \nonumber \\
& =& E_{m0}^{\downarrow \uparrow } -\tilde{V}^{(1)}_{m00m} (q)
\; , \label{eq27}  \\
\omega_m -m \bar{\omega}_c +g\mu_B B &=& \Delta
E_m^+  \nonumber \\
& =& E_{m0}^{\uparrow \downarrow} -\tilde{V}^{(1)}_{m00m} (q)
\; . \label{eq28}  
\end{eqnarray}
Here $\Delta E_m^+ = \Delta E_m^-$ as $E_{m0}^{\downarrow \uparrow
} = E_{m0}^{\uparrow \downarrow}$ due to the equal population of
both the spins. The change in energy due to the Coulomb
interaction for the two lowest modes ($m=1,2$) are given by
\begin{eqnarray}
\Delta E_1^{\pm} (q) &=& 
\frac{e^2}{\epsilon l_0} \frac{1}{2} \sqrt{\frac{\pi}{2}} 
\left\{ 1- e^{- \bar{{\bf q}}^2 /2}
\left[ (1+ \bar{{\bf q}}^2) I_0 \left( \frac{\bar{{\bf
q}}^2}{2} \right) - \bar{{\bf q}}^2 I_1 \left(
\frac{\bar{{\bf q}}^2}{2} \right) \right] \right\} 
\; , \label{eq29} \\
\Delta E_2^{\pm} (q) &=& \frac{e^2}{\epsilon l_0} \frac{1}{8}
\sqrt{\frac{\pi}{2}} \left\{ 5- e^{- \bar{{\bf q}}^2 /2}
\left[ (3+ 2\bar{{\bf q}}^2 + 2 \bar{{\bf q}}^4)
I_0 \left( \frac{\bar{{\bf
q}}^2}{2} \right) \right. \right. \nonumber \\
& & \;\;\;\;\;\;\; \left. \left. - (4\bar{{\bf q}}^2 +2 \bar{{\bf
q}}^4) I_1 \left(
\frac{\bar{{\bf q}}^2}{2} \right) \right] \right\} 
\label{eq30}
\end{eqnarray}
respectively. These are shown in Fig.~2.
Note that the Coulomb interaction
does not contribute to 
the energy of $m=1$ mode at $ql_0 = 0$. 
The unpolarized states with $\nu = 2/(4s\pm
1)$, {\it i.e.}, the states with 2 as a numerator such as $2,2/3,2/5$
have similar spin flip excitations as above since they all have
same $l_0$.

Case-III: Consider the simplest case $\vert p_ \uparrow \vert
=2$ and $\vert p_ \downarrow \vert =1$ as an example of
partially polarized QHS. The dispersion relation corresponding
to $\delta S_z =+1$ excitations is given by
\begin{equation}
\omega_m -m \bar{\omega}_c +g\mu_B B = \Delta
E_m^+  = E_{m0}^{\uparrow \downarrow } -\tilde{V}^{(1)}_{m00m} (q)
\label{eq31}
\end{equation}
with $m>1$. Therefore there is no mode at $\bar{\omega}_c
-g\mu_B B$ in this case. $\delta S_z =-1$ type of
spin flip excitations may occur from any of the two filled LL by
up spins. They have two classes of dispersion relations given by
\begin{eqnarray}
\omega_m -m \bar{\omega}_c -g\mu_B B &=& \Delta
E_m^{(1)-}  = E_{m0}^{\downarrow \uparrow } -\tilde{V}^{(1)}_{m00m} (q)
\;  ;\; m>0  \; , \label{eq32}  \\
\omega_m -(m-1)\bar{\omega}_c -g\mu_B B &=& \Delta
E_m^{(2)-}  = E_{m1}^{\downarrow \uparrow } -\tilde{V}^{(1)}_{m11m} (q)
\;  ;\; m>1  \; . \label{eq33}  
\end{eqnarray}
The energies $\Delta E_1^{(1)-} (q)$, $\Delta E_2^{(1)-}(q)$,
and $\Delta E_2^{(2)-} (q)$ are shown in Fig.~3.
Figure 4 represents
$\Delta E_1^+ (q)$. The states with $\nu =3/(6s\pm
1)$, {\it i.e.}, all the states with numerator 3 such as 3, 3/5, 3/7
have the similar spin-flip excitations as shown above.

\subsection{Spin-wave excitations}

When the Fermi energy lies between two spin split levels in the same
LL, the particles may be excited within the same LL by flipping
the spin. This excitation is possible only in a ferromagnetic
ground state. Therefore, fully polarized and partially polarized
QHS are possible candidates to observe this kind of excitations.
We shall see below that the Coulomb interactions do not change
the spin-wave excitation energy at $q=0$, as required by Larmor's
theorem. The long wave length spin wave becomes gapless at
$g=0$.

{\it Fully polarized states}: 
Consider the fully polarized $(p_ \uparrow =1, \, p_ \downarrow
=0)$ QHS first. The spin wave dispersion relation is obtained as
\begin{equation}
\omega -g\mu_B B = \Delta E (q)= E_{00}^{\downarrow
\uparrow} - \tilde{V}^{(1)}_{0000}(q) \; . \label{eq34}
\end{equation}
The explicit form of $\Delta E (q)$ is given by
\begin{equation}
\Delta E(q) = \frac{e^2}{\epsilon l_0} \sqrt{\frac{\pi}{2}}
\left[ 1- e^{- \bar{{\bf q}}^2/2} I_0 \left(
\frac{\bar{{\bf q}}^2}{2} \right) \right]  \; .
\label{eq35}
\end{equation}
The dispersion energy $\Delta E (q)$ is shown in Fig.~5. The
states $\nu =1$ and $\nu =1/(2s+1)$ follow the above dispersion
relation in their spin wave excitations. 
At very low $ql_0$, the spin
wave is quadratically dispersed as
\begin{equation}
\omega = g \mu_B B + \frac{e^2}{\epsilon l_0}
\frac{1}{2} \sqrt{\frac{\pi}{2}} (ql_0)^2 \; .
\label{eq36}
\end{equation}
Nakajima and Aoki \cite{naka:94} also have numerically found spin wave
excitations for these Laughlin states using CF picture. They 
used a reduced Haldane pseudo-potential \cite{hald:87}
for CF's to carry out the
computation in a spherical geometry.
The energy cost to excite a spin down quasiparticle and a spin
up quasihole is $\sqrt{\frac{\pi}{2}} \frac{e^2}{\epsilon l_0}$. This
excitation corresponds to $\bar{{\bf q}}^2 \rightarrow  \infty $
in Eq.~(\ref{eq35}).

{\it Partially polarized states}:
For the simplest case of partially polarized QHS, $\vert p_
\uparrow \vert =2$ and $\vert p_ \downarrow \vert =1$. In this
case, the spin wave dispersion relation is given by
\begin{equation}
\omega -g\mu_B B = \Delta E = E_{11}^{\downarrow
\uparrow} - \tilde{V}^{(1)}_{1111}(q) \; . \label{eq37}
\end{equation}
We find the dispersion energy to be
\begin{equation}
\Delta E(q) = \frac{e^2}{\epsilon l_0 }\frac{1}{4} 
\sqrt{\frac{\pi}{2}}
\left[ 3- e^{- \bar{{\bf q}}^2/2} \left\{ (3- 2
\bar{{\bf q}}^2 +2 \bar{{\bf q}}^4) I_0 \left(
\frac{\bar{{\bf q}}^2}{2} \right) -2 \bar{{\bf q}}^2
I_1 \left( \frac{\bar{{\bf q}}^2}{2} \right) \right\} 
\right] 
\label{eq38}
\end{equation}
which is shown in Fig.~6.
The spin wave excitations of the states
 with $\nu =3$, and $\nu =3/(6s\pm 1)$ such as 3, 3/5, 3/7 have
 the above dispersion relation. At very low $ql_0$, this is
 again quadratically dispersed. The dispersion relation then is
 given by
\begin{equation}
\omega = g \mu_B B + \frac{e^2}{\epsilon l_0}
\frac{7}{16} \sqrt{\frac{\pi}{2}} (ql_0)^2 \; .
\label{eq39}
\end{equation}
In this case, the energy necessary to create a spin up
quasiparticle and a spin down quasihole pair is $\frac{3}{4}
\sqrt{\frac{\pi}{2}} \frac{e^2}{\epsilon l_0}$ (see
Eq.~\ref{eq38}).
Similarly other partially polarized QHS such as $\nu =5$, 5/9,
5/11 also possess spin wave excitations.

\section{Summary and discussions}

We have studied spin flip and spin wave excitations for arbitrarily
polarized quantum Hall states in the time dependent Hartree-Fock
approximation. We have employed fermionic Chern-Simons theory within 
the composite fermion picture. The spin flip correlation functions
do not get renormalized by the fluctuations of Chern-Simons gauge
field over its mean value which describes the ground state. That is 
they depend entirely on $p_\uparrow + p_\downarrow $ and the 
excitations are hence identical to the excitations in the
corresponding integer state, same for the replacement 
magnetic length $l \rightarrow l_0 $.
We have shown that the Coulomb interaction between composite fermions
produces an additional gap for spin flip excitations, in general.
On the other hand, as a consequence of Larmor's theorem, neutral
spin wave modes are gapless as $g \rightarrow 0$.
There persists gap for creating a quasiparticle and quasihole pair.
However, the lowest energy charged excitations in the ferromagnetic
ground state of $\nu = 1/(2s+1)$ are skyrmions 
\cite{sondhi:93} which are experimentally verified 
\cite{barret:95,scheml:95,gberg:96} at $\nu =1$. It has been 
found by Wu and Sondhi \cite{sondhi:95} that the odd integer states
also possess skyrmion like excitations although they have higher
energies compared to quasiparticle and quasihole pair
excitations. It is now interesting to observe the occurrence of 
skyrmions in all the arbitrarily polarized quantum Hall states by
a general frame work. An attempt in this line is in progress.

\section*{Acknowledgments}

   I thank the anonymous referee of the paper in Ref.~\cite{ssm:96b}
for suggesting the problem
which I have studied here. Many discussions with 
V. Ravishankar are gratefully acknowledged.
Finally, I thank JNCASR, Bangalore, for financial assistance.

\appendix

\section{Response function in TDHFA}

In this appendix, we shall evaluate $\chi^{ab}$ in TDHFA. We
adopt the diagrammatic method that was developed by Kallin and
Halperin ~\cite{kallin:84}.

The single particle Hartree-Fock Green's function is given by
\begin{equation}
G_n^r (\omega) = \frac{1}{\omega - \epsilon_n^r -\Sigma_n^r
+i\delta (n,r)} \; , 
\label{eqb1}
\end{equation}
where $\delta (n,r) = 0^+ $ for $n <p_r $ and $\delta (n,r) =
0^-$ for $n \geq p_r$.  
$G_n^r $ is diagrammatically shown in Fig.~7.
In the strong field
approximation (the effective field $\bar{B}$ here),
the self energy, which is diagrammatically shown in Fig.~7, is given by
\begin{equation}
\Sigma_n^r =
-\int \frac{d^2r}{2\pi l_0^2} V(r)
e^{-r^2/2l_0^2} L^1_{p_r-1} \left( \frac{r^2}{2l_0^2} \right)
L_n^0 \left( \frac{r^2}{2l_0^2} \right) \; ,
\label{eqb2}
\end{equation}
where $V(r) =e^2 /\epsilon r$ is the Coulomb potential. Note
that the same index $r$ stands for both spin and spatial
coordinates. The associated Laguerre polynomial is given by
\begin{equation}
L_n^m (x) = \frac{1}{n!}e^x x^{-m} \frac{d^n}{dx^n} \left(
e^{-x} x^{n+m} \right) \; .
\label{eqb3}
\end{equation}

The response function $\chi^{ab} (\omega \, {\bf q})$ may be expressed as 
(see Fig.8(a))
\begin{equation}
\chi^{ab} (\omega ,\, {\bf q}) = \frac{e^2}{2\pi l_0^2}
\sum_{n_1,n_2}\sum_{r_1,r_2}
M_{n_1n_2} ({\bf q}) D_{n_1n_2}^{r_1r_2} (\omega)
\Gamma_{n_1n_2}^{r_1r_2} (\omega ,\, {\bf q}) 
\langle r_1 \left\vert \sigma_a \right\vert r_2 \rangle 
\langle r_2 \left\vert \sigma_b \right\vert r_1 \rangle  \; .
\label{eqb4}
\end{equation}
Here the matrix element (contribution of the left vertex of Fig.~8(a))
is obtained as 
\begin{equation}
M_{n_1n_2} ({\bf q}) = \left( \frac{2^{n_2}n_2! }{2^{n_1}n_1!}
\right)^{1/2} e^{-\bar{{\bf q}}^2/2} \left[ (q_x+iq_y)l_0
\right]^{n_1-n_2} L_{n_2}^{n_1-n_2} (\bar{{\bf q}}^2) \; .
\label{eqb5}
\end{equation}
The two-particle propagator (see Fig.~8(a))
is given by
\begin{eqnarray} 
& & D_{n_1n_2}^{r_1r_2} (\omega)  = \int \frac{d\omega^\prime}{2\pi i}
G_{n_1}^{r_1} (\omega^\prime)G_{n_2}^{r_2} (\omega +\omega^\prime
)  \label{eqb6} \\
&=& \left[ \frac{\theta (p_{r_1}-n_1)\theta (n_2-p_{r_2} -1)}{ \omega
-(\epsilon_{n_1}^{r_1} -\epsilon_{n_2}^{r_2}) 
-(\Sigma_{n_1}^{r_1} -\Sigma_{n_2}^{r_2}) 
+i\eta} 
- \frac{\theta (p_{r_2}-n_2)\theta (n_1-p_{r_1} -1)}{ \omega
-(\epsilon_{n_1}^{r_1} -\epsilon_{n_2}^{r_2}) 
-(\Sigma_{n_1}^{r_1} -\Sigma_{n_2}^{r_2}) 
-i\eta}  \right]  \, ,  \nonumber \\
& & \label{eqb7}
\end{eqnarray}
where the function $\theta (x)$ is defined as
\begin{equation}
\theta (x) \equiv \left\{ \begin{array}{lll}
1 & {\rm for}& x>0 \\
0 & {\rm for}& x<0 \end{array} \right. \; .
\label{eqb8}
\end{equation}
Finally, $\Gamma_{n_1n_2}^{r_1r_2}$ which is the corrected
vertex (due to Coulomb interaction between the particle and hole
pair) function is diagrammatically shown in Fig.~8(b). Here we have
assumed that only a single exciton is present at a time. In other
words, $e^2 /\epsilon l_0 \ll \bar{\omega}_c$.

Now if we define
\begin{equation}
\Phi_{n_1n_2}^{r_1r_2} (\omega ,\, {\bf q}) \equiv 
D_{n_1n_2}^{r_1r_2} (\omega) 
\Gamma_{n_1n_2}^{r_1r_2} (\omega ,\, {\bf q}) \; ,
\label{eqb9}
\end{equation}
then the response function can be written as
\begin{equation}
\chi^{ab} (\omega ,\, {\bf q}) = \frac{e^2}{2\pi l_0^2}
\sum_{r_1,r_2}\sum_{n_1,n_2}
M_{n_1n_2}({\bf q}) \Phi_{n_1n_2}^{r_1r_2} (\omega ,\, {\bf q})
\langle r_1 \left\vert \sigma_a \right\vert r_2 \rangle 
\langle r_2 \left\vert \sigma_b \right\vert r_1 \rangle  \; ,
\label{eqb10}
\end{equation}
where $\Phi_{n_1n_2}^{r_1r_2} (\omega ,\, {\bf q})$ satisfies the
matrix equation (independent of spin indices),
\begin{equation}
\sum_{n_3,n_4}\sum_{r_3,r_4} \left[ \delta_{n_1,n_3}
\delta_{n_2,n_4}\delta_{r_1,r_3}\delta_{r_2,r_4} \left[ \left\{
D (\omega) \right\}^{-1} \right]^{r_3r_4}_{n_3n_4} -
\delta_{r_1,r_3}\delta_{r_2,r_4} \tilde{V}^{(1)}_{n_1 n_4
n_2n_3} ({\bf q}) \right]  
\Phi_{n_3 n_4}^{r_3 r_4} 
= M^\ast_{n_1n_2} ({\bf q}) \; .
\label{eqb11}
\end{equation}
Diagrammatically, $M_{n_1n_2}^\ast ({\bf q})$ is given by the
first diagram of Fig.~8(b). The interaction between an
excited electron and hole is represented by the ladder
diagrams in the second diagram of Fig.~8(b). The
corresponding matrix element due to to Coulomb interaction is
given by
\begin{eqnarray}
\tilde{V}^{(1)}_{n_1n_4n_2n_3} ({\bf q}) 
&=& \left( \frac{2^{n_4}2^{n_2} n_4!
n_2!}{2^{n_1}2^{n_3}n_1!n_3!} \right)^{1/2} \int
\frac{d^2r}{2\pi l_0^2} V({\bf r} -l_0^2{\bf q} \times \hat{z})
e^{-r^2/2l_0^2} \nonumber \\
&\times & \left[ \frac{x+iy}{l_0}\right]^{n_1-n_2} 
 \left[ \frac{x-iy}{l_0}\right]^{n_3-n_4} 
 L_{n_2}^{n_1-n_2} \left( \frac{r^2}{2l_0^2} \right)
 L_{n_4}^{n_3-n_4} \left( \frac{r^2}{2l_0^2} \right) \, ,
 \nonumber \\
 & & \label{eqb12}
 \end{eqnarray}
for $n_2 \leq n_1$ and $n_4 \leq n_3 $. If $n_2 > n_1$, then $n_2 $ and 
$n_1$ in the right side of Eq.~(\ref{eqb12}) are interchanged; and
similarly if $n_4 >n_3$, then $n_3$ and $n_4$ are interchanged.

Solving $\Phi_{n_1n_2}^{r_1r_2} (\omega ,\, {\bf q})$ from
Eq.~(\ref{eqb11}), we substitute it in Eq.~(\ref{eqb10}) to
obtain the response function,
\begin{eqnarray}
& & \chi^{ab} (\omega ,\, {\bf q}) = \frac{e^2}{2\pi l_0^2} 
\sum_{r_1}\sum_{r_2}  
\langle r_1 \left\vert \sigma_a \right\vert r_2 \rangle 
\langle r_2 \left\vert \sigma_b \right\vert r_1 \rangle  \nonumber \\ 
&\times & \left[ \sum_{n_2<p_{r_1}} \sum_{n_1 \geq p_{r_2}}
\frac{ \left\vert M_{n_1n_2} ({\bf q})\right\vert^2 }{
\omega - (\epsilon_{n_1}^{r_2} -\epsilon_{n_2}^{r_1})
- (\Sigma_{n_1}^{r_2} -\Sigma_{n_2}^{r_1})
+\tilde{V}^{(1)}_{n_1n_2n_2n_1}({\bf q}) +i \eta } \right.
\nonumber \\
& & - \left. \sum_{n_2<p_{r_2}} \sum_{n_1 \geq p_{r_1}}
\frac{ \left\vert M_{n_1n_2} ({\bf q})\right\vert^2 }{
\omega + (\epsilon_{n_1}^{r_1} -\epsilon_{n_2}^{r_2})
+ (\Sigma_{n_1}^{r_1} -\Sigma_{n_2}^{r_2})
-\tilde{V}^{(1)}_{n_1n_2n_2n_1}({\bf q}) -i \eta } \right]
\, , \nonumber \\
& & \label{eqb16}
\end{eqnarray}
where $\sigma_a$ $(\sigma_b)$ creates spin state $r_1$ $(r_2)$
destroying spin state $r_2$ $(r_1)$. Therefore, $r_1 \neq r_2$.
Note that the bubble diagrams (see the third diagram of Fig.~8(b)) 
do not contribute to the spin-flip
response function. This is because in this response function,
particle and hole possess different spins, and the Coulomb
interaction can not flip the spin.

\newpage 

\begin{center}
{\bf FIGURE CAPTIONS}
\end{center}

\bigskip

\noindent
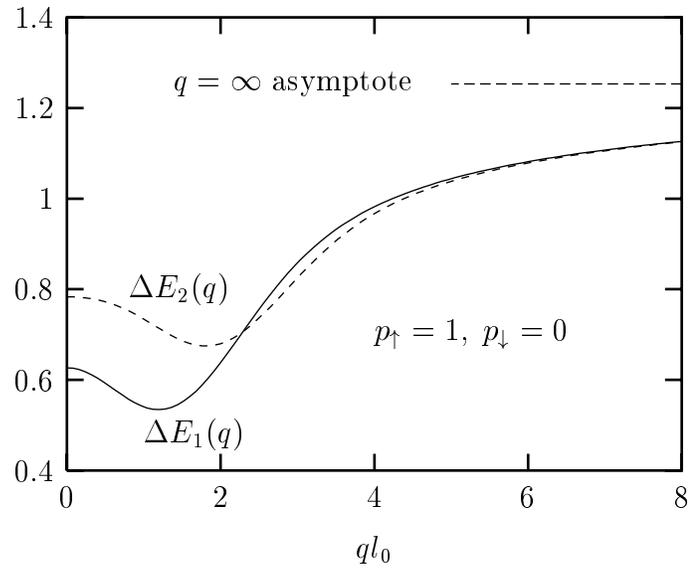
FIG.~1. $\Delta E_1 (q)$ and $\Delta E_2 (q)$ are plotted in the
units of $e^2 / \epsilon l_0$ against $ql_0$. $q=\infty$
asymptotes are same for both the cases.

\medskip

\noindent
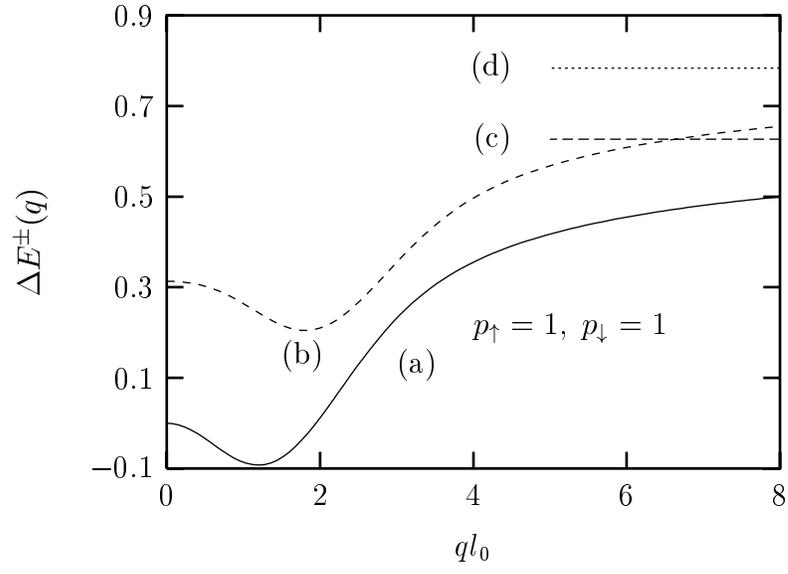
FIG.~2. (a) $\Delta E_1^\pm (q)$ and (b) $\Delta E_2^\pm (q)$
are plotted in the units of $e^2 / \epsilon l_0$ against $ql_0$.
The lines (c) and (d) are the respective $q=\infty$ asymptotes.

\medskip

\noindent
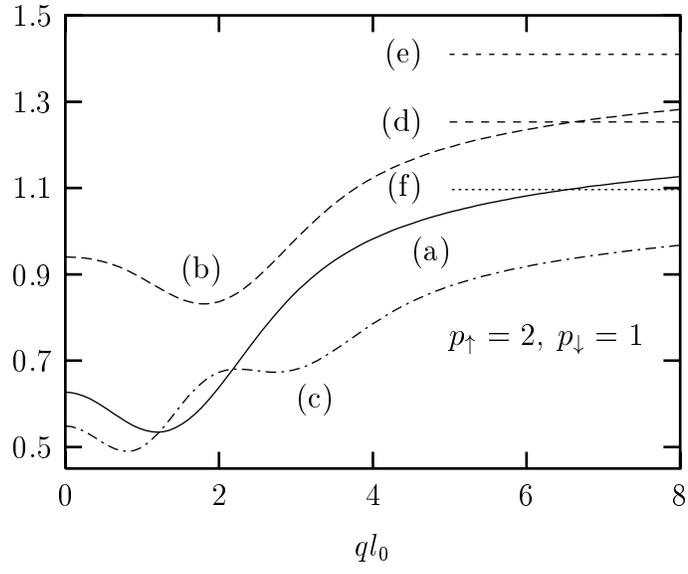
 FIG.~3. (a) $\Delta E_1^{(1)-}(q)$, (b) $\Delta E_2^{(1)-}
(q)$, and (c) $\Delta E_1^{(2)-} (q)$ are plotted in the units
of $e^2 / \epsilon l_0$ against $ql_0$. The lines (d), (e), and
(f) are the respective $q =\infty$ asymptotes.

\medskip

\noindent
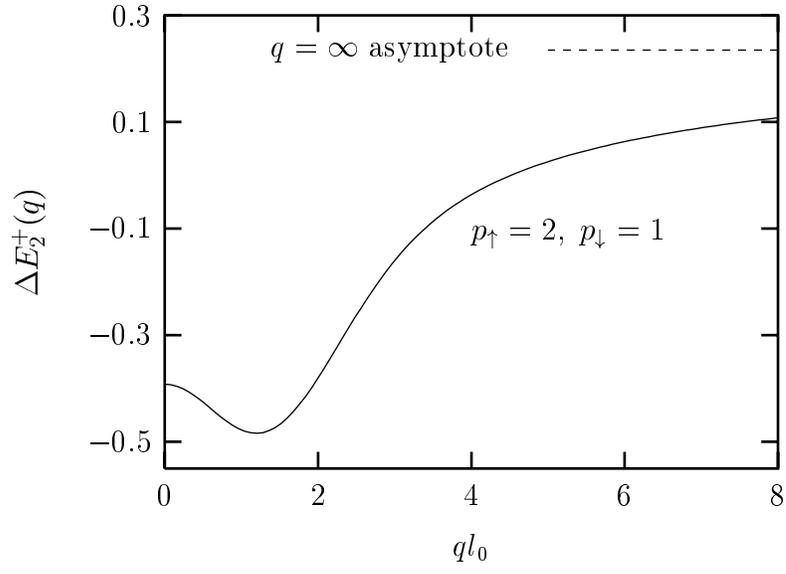
FIG.~4. The spin flip mode, corresponding to flipping the spin
from down to up, $\Delta E_1^+ (q)$ is shown in the unit of $e^2
/ \epsilon l_0$.

\medskip

\noindent
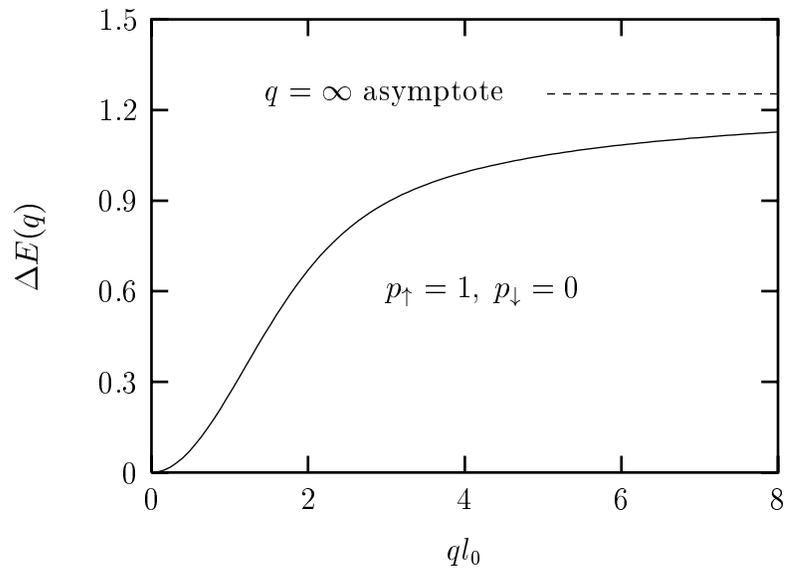
FIG.~5. The spin-wave mode is shown in the unit of $e^2 /
\epsilon l_0$ for $p_ \uparrow  =1,\, p_
\downarrow =0$, with the omission of Zeeman energy.

\medskip

\noindent
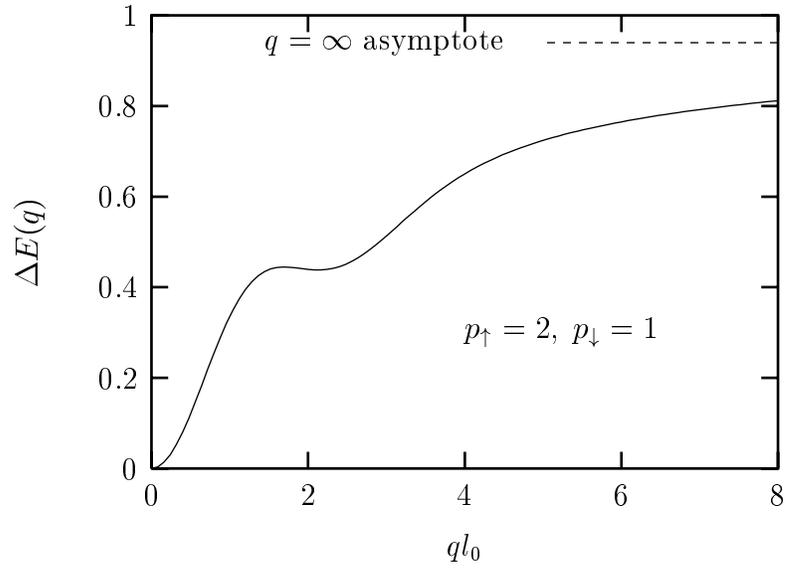
FIG.~6. The spin-wave mode is shown in the unit of $e^2 /
\epsilon l_0$ for $p_ \uparrow  =2,\, p_
\downarrow =1$, with the omission of Zeeman energy.

\medskip

\noindent
FIG.~7. Thick (thin) lines represent single
particle Hartree-Fock (bare) Green's functions. 
The wiggly lines represent the Coulomb interaction. (a) The self
energy $\Sigma_n^r$ for the particles of spin index $r$ in the
$n$ th Landau level. (b) The single particle Hartree-Fock
Green's function $G_n^r$ for $n$ th landau level and spin index $r$.

\medskip

\noindent
FIG.~8. $ \alpha_i = (n_i ,\, r_i)$ denotes collectively
the Landau level index $n_i$ and the spin index $r_i$. Thick (thin)
lines represent single particle Hartree-Fock (bare)
Green's functions. The dashed lines 
represent the probes. (a) Response function
$\chi^{ab} (\omega , \, {\bf q})$ is diagrammatically shown. The
shaded portion represents the vertex correction due to the
Coulomb interaction. (b) The vertex function
$\Gamma_{n_1n_2}^{r_1r_2} (\omega , \, {\bf q})$ is
diagrammatically shown.


\begin{references}
\bibitem[*]{email} Electronic address: ssman@physics.iisc.ernet.in
\bibitem{jain:89} J. K. Jain, Phys. Rev. Lett. {\bf 63},
                  199 (1989).
\bibitem{lopez:91} A. Lopez and E. Fradkin, 
            Phys. Rev. {\bf B 44}, 5246 (1991). 
\bibitem{halp:93} B. I. Halperin, P. A. Lee, and N. Read,
		Phys. Rev. {\bf B 47}, 7312 (1993).
\bibitem{du:93} R. R. Du, H. L. Stormer, D. C. Tsui,
			  L. N. Pfeiffer, and K. W. West,
            Phys. Rev. Lett. {\bf 70}, 29444 (1993).
\bibitem{will:93} R. L. Willett, R. R. Ruel, M. A. Paalanen, K.
		W. West, and L. N. Pfeiffer,
		Phys. Rev. {\bf B 47}, 7344 (1993);
		{\it ibid.} {\bf 71}, 3846 (1993).
\bibitem{kang:93} W. Kang, H. L. Stormer, L. N. Pfeiffer,
                      K. W. Baldwin, and K. W. West, 
            Phys. Rev. Lett. {\bf 71}, 3850 (1993).
\bibitem{lead:94} D. R. Leadley, H. L. Stormer, C. T. Foxon, and
		J. J. Harris, Phys. Rev. Lett. {\bf 72}, 1906
		(1994).
\bibitem{du:94} R. R. Du, H. L. Stormer, D. C. Tsui,
			  L. N. Pfeiffer, and K. W. West,
            Phys. Rev. Lett. {\bf 73}, 3274 (1994).
\bibitem{mano:94} H. C. Manoharan, M. Shayegan, and S. J. Klepper,
            Phys. Rev. Lett. {\bf 73}, 3270 (1994).
\bibitem{gold:94} V. J. Goldman, B. Su, and J. K. Jain, 
		Phys. Rev. Lett. {\bf 72}, 2065 (1994).
\bibitem{ying:94} X. Ying, V. Bayot, M. B. Santos, and
		M. Shayegan, Phys. Rev. {\bf B 50}, 4969 (1994).
\bibitem{kuku:94} I. V. Kukushkin, R. J. Haug, K. v. Klitzing,
			   and K. Ploog,
        Phys. Rev. Lett. {\bf 72}, 736 (1994).			    
\bibitem{bayot:95} V. Bayot, E. Grivei, H. C. Manoharan, X.
		Ying, and M. Shayegan, Phys. Rev. {\bf B 52}, R8621
		(1995).
\bibitem{kuku:95} I. V. Kukushkin, R. J. Haug, K. v. Klitzing,
			   and K. Ploog,
            Phys. Rev. {\bf B 51}, 18045 (1995).
\bibitem{ssm:96a} S. S. Mandal and V. Ravishankar, Phys. Rev. {\bf
         B 54}, 8688 (1996). 
\bibitem{ssm:96b} S. S. Mandal and V. Ravishankar, Phys. Rev. {\bf
         B 54}, 8699 (1996). 
\bibitem{stein:83} D. Stein, K. v. Klitzing, and G. Weimann, Phys.
        Rev. Lett. {\bf 51}, 130 (1983).
\bibitem{kallin:84} C. Kallin and B. I. Halperin, Phys. Rev. {\bf 
        B 30}, 5655 (1984).
\bibitem{longo:93} J. P. Longo and C. Kallin, Phys. Rev. {\bf B
        47}, 4429 (1993).
\bibitem{fnote} In Ref.~\cite{ssm:96a}, the Zeeman term includes 
        CS gauge fields as well. Strictly speaking, the $g$ factor
        then should have been renormalized one. Here the $g$
        factor is the bare value of the system. Thus only applied
        magnetic field is included in the Zeeman term.
\bibitem{naka:94} T. Nakajima and H. Aoki, Phys. Rev. Lett. {\bf 73},
        3568 (1994).
\bibitem{hald:87} F. D. M. Haldane, in {\it The Quantum Hall Effect},
       edited by R. E. Prange and S. M. Girvin, (Springer-Verlag,
        New York), 1987.
\bibitem{sondhi:93} S. L. Sondhi, A. Karlhede,
			   S. A. Kivelson, and E. H. Rezayi,
            Phys. Rev. {\bf B 47}, 16419 (1993).
\bibitem{barret:95} S. E. Barrett, G. Dabbagh, L. N. Pfeiffer,
		   K. W. West, and R. Tyco, Phys. Rev. Lett. {\bf
		   74}, 5112 (1995).
\bibitem{scheml:95} A. Schmeller, J. P. Eisenstein, L. N.
		 Pfeiffer, and K. W. West, Phys. Rev. Lett. {\bf 
		 75}, 4290 (1995).
\bibitem{gberg:96} E. H. Aifer, B. B. Goldberg, and D. A.
		   Broido, Phys. Rev. Lett. {\bf 76}, 680 (1996).
\bibitem{sondhi:95} X. G. Wu and S. L. Sondhi, 
		     Phys. Rev. {\bf B 51}, 14725 (1995).

\end{references}
\end{document}